\documentclass[10pt, letterpaper, conference]{IEEEtran}
\usepackage{amsmath,amssymb,amsfonts}
\usepackage[ruled,vlined]{algorithm2e}
\usepackage{booktabs}
\usepackage{pifont}
\usepackage{soul}
\usepackage{tabularx}
\usepackage{adjustbox}
\usepackage{array}
\usepackage{xparse}
\usepackage[hidelinks]{hyperref}
\usepackage{etoolbox}
\usepackage{wrapfig}
\usepackage{xcolor}
\usepackage{wasysym}
\usepackage{xcolor}
\usepackage{listings}
\usepackage{multirow}
\usepackage{makecell}
\usepackage{tablefootnote}
\usepackage{comment}
\usepackage[numbers]{natbib}
\usepackage{graphicx}
\usepackage{textcomp}
\usepackage{url}
\usepackage[nolist, printonlyused]{acronym}
\usepackage{mathtools}
\usepackage[utf8]{inputenc}
\usepackage{balance}
\usepackage{fancyhdr}

\usepackage{tikz}
\usetikzlibrary{backgrounds}
\usetikzlibrary{positioning,fit,patterns}
\usetikzlibrary{shapes} 
\usetikzlibrary{arrows}
\usetikzlibrary{arrows.meta}
\usetikzlibrary{positioning,automata}
\usetikzlibrary{calc}

\newcommand{\cmark}{\raisebox{-0.15em}{$\checked$}}
\newcommand{\Cmark}{\raisebox{-0.15em}{\ding{51}}}
\newcommand{\xmark}{\raisebox{-0.15em}{\hspace{0.1em}$\times$}}
\newcommand{\ndmark}{$\ddagger$}
\newcommand{\ignore}[1]{}

\NewDocumentCommand{\irregularline}{%
  O     {2mm}   % Amplitude of irregularity. Optional. Default value = 2mm
  m             % First point
  m             % Second point
  D   <> {20}   % Number of peaks. Optional. Default value = 20
}{{%
  \coordinate (old) at #2;
  \foreach \i in {1,2,...,#4}{
    \draw (old) -- ($ ($#2!\i/(#4+1)!#3$) + (0,#1*rand) $) coordinate (old);
  }
  \draw (old) -- #3;
}}

\newcommand{\circlenumber}[1]{\begin{tikzpicture}[baseline=(char.base)]
\node(char)[state,inner sep=1pt, minimum size=0pt]{#1};
\end{tikzpicture}}

\newcommand\copyrighttext{%
  \footnotesize \textcopyright 2021 IEEE.  Personal use of this material is permitted.  Permission from IEEE must be obtained for all other uses, in any current or future media, including reprinting/republishing this material for advertising or promotional purposes, creating new collective works, for resale or redistribution to servers or lists, or reuse of any copyrighted component of this work in other works.}
\newcommand\copyrightnotice{%
\begin{tikzpicture}[remember picture,overlay]
\node[anchor=south,yshift=10pt] at (current page.south) {\fbox{\parbox{\dimexpr\textwidth-\fboxsep-\fboxrule\relax}{\copyrighttext}}};
\end{tikzpicture}%
}

\title{The Master and Parasite Attack}
\author{\IEEEauthorblockN{Lukas Baumann, Elias Heftrig, Haya Shulman and Michael Waidner}
\IEEEauthorblockA{
Fraunhofer Institute for Secure Information Technology \\
Darmstadt, Germany\\
}
}

\begin{document}

\maketitle
\copyrightnotice
 
\begin{abstract}
We explore a new type of malicious script attacks: the {\em persistent parasite attack}. Persistent parasites are stealthy scripts, which persist for a long time in the browser's cache. We show to infect the caches of victims with parasite scripts via TCP injection.
    
Once the cache is infected, we implement methodologies for propagation of the parasites to other popular domains on the victim client as well as to other caches on the network. We show how to design the parasites so that they stay long time in the victim's cache not restricted to the duration of the user's visit to the web site. We develop covert channels for communication between the attacker and the parasites, which allows the attacker to control which scripts are executed and when, and to exfiltrate private information to the attacker, such as cookies and passwords. We then demonstrate how to leverage the parasites to perform sophisticated attacks, and evaluate the attacks against a range of applications and security mechanisms on popular browsers. Finally we provide recommendations for countermeasures.
\end{abstract}

\section{Introduction}
\label{sec:introduction}

Malware is the most significant threat to computer users; and, from all types of malware, the greatest threat are {\em advanced persistent threats (APTs)}, which may reside in the victim's computer for long time without detection. However, installing malware - and APTs - on a victim's computer can be challenging. Either the attacker has to trick the user into installing a malicious software, or a vulnerability in user's system or applications must be exploited to install the malicious software without the user's permission. 

In contrast, browsers automatically download and execute code from remote websites. Unlike with malicious software, such code runs in a restricted {\em sandbox}, rather than the `native' code of malware. The goal of the sandbox is to prevent attacks by automatically downloaded code. Furthermore, remotely-downloaded Javascript code executes only while the browser visits the relevant website, i.e., {\em ephemerally} rather than {\em persistently}, and stays in the browser for as long as there is place in the cache and the website does not serve a new version of the same object. Hence, any malicious computation, such as abusing users' compute power for crypto-currency mining \cite{papadopoulos2018master}, can affect the users only as long as they visit the website or until the script is evicted from the cache. 

 In summary, Javascript and other remote code executed in a sandbox is believed to be a relatively minor -- in contrast to user or root level malware -- ephemeral (non-persistent) threat. However, in this work, we show that remotely-downloaded Javascript can be {\em persistent} and poses a serious threat. In particular, we show that attackers can perform an attack so that these scripts are executed permanently in victims' browsers and then actively spread this attack to other domains. The first step of our attack includes injecting and spreading a malicious script, which we call {\em parasite}. We then show how these scripts can be abused by the attacker to perform various attacks against applications, such as stealing clients' credentials on popular web sites or bypassing a two factor authentication. We evaluate the effectiveness of these attacks on popular browsers. Typically such attacks should not be possible: a script coming from an attacker would be restricted by the Same Origin Policy (SOP) and hence should not be able to access data of objects coming from other domains. {\em To bypass the SOP the attacker camouflages the malicious (parasite) script to appear as if it originated from a real origin website.} We develop methodologies to keep the parasite in the victim's browser over long time periods and to control it to launch attacks against different applications, even after the client moves to a different (e.g., home) network. The parasite uses modules we developed for {\em Command and Control (C\&C)} to the attacker and for {\em propagation}. We demonstrate attacks that we implemented using the parasites and provide results of our experimental evaluation.

{\bf Attack Overview.} The attack consists of four modules: (1) cache eviction, (2) injection into transport layer, (3) parasite construction and finally (4) applying the parasites to launch sophisticated attacks. 

As a first step, the attacker performs cache eviction (Section \ref{sec:http-eviction}), to remove cached objects of popular target domains, then causes the browser to issue an HTTP request to the target website. 
The attacker injects a spoofed TCP segment with the infected object into the HTTP response from the server, which delivers the malicious payload into victim's cache (Section \ref{sc:inject:transport}). The injection can be performed with a remote attacker by launching DNS cache poisoning \cite{shulman2014fragmentation,cns:frag:dns,shulman2015towards,herzberg2013socket,gilad2013off,brandt2018domain,herzberg2012security,klein2017internet,man2020dns,alharbi2019collaborative,zheng2020poison,brandt2018domain}. 
As soon as the parasite is cached in the victim's browser, it starts infecting other domains in the cache and establishes a C\&C to the remote attacker to carry out advanced attacks against applications (Section \ref{sec:applications}).

{\bf Contributions.} Our goal is to explore the attacks against applications that remotely controlled parasite scripts can launch and the feasibility of constructing such parasite scripts in the wild.
Our contributions can be summarised as follows:

$\bullet$ We identify websites which use persistent objects. We measure the persistency and prevalence of these objects on popular websites. We found that more than $87\%$ of the websites have at least one object persistent over a period of 5 days. We modify these objects attaching to them malicious parasites.

$\bullet$ We develop methodologies to force the browsers to keep our parasite scripts in the cache even after the victim stops visiting the website whose object was infected with a parasite, and even after the victim moves to a different network. 

$\bullet$ We devise techniques to bypass SOP, allowing the parasites to propagate between different domains and caches. Our techniques can be systematically launched via TCP injection in contrast to previous work which exploited bugs to bypass SOP \cite{qian2012off}. Since our approach leverages the standard behaviour of the systems it is much more difficult to block.

$\bullet$ We develop and evaluate command and control channels for communication between the attacker and the parasites.

$\bullet$ We developed a taxonomy of popular caches and evaluate our attacks experimentally against them. These are caches that can be infected with our parasite and then used to attack the applications.

$\bullet$ We developed a taxonomy of the exploits against popular applications which we attacked with our parasites botnet. The applications range from financial and banking systems to hardware components on devices (such as camera and mic).

{\bf Organisation.} In Related Work, Section \ref{sec:works} we review research relevant to our work. In Section \ref{sc:attack} we explain our attacker model. In Sections \ref{sec:http-eviction}, \ref{sc:inject:transport} and \ref{sc:parasite} we present the implementation of our attacker and report results of our evaluations in the Internet. In Section \ref{sec:applications} we demonstrate how parasites can be leveraged to attack a wide range of systems and applications and provide recommendations for countermeasures in Section \ref{sec:countermeasures}. We conclude this work in Section \ref{sec:conclusions}.

\vspace{-10pt}
\section{Related Work}
\label{sec:works}

\textbf{Malicious JavaScript attacks}. Due to policies like SOP \cite{ruderman2009same}, CORS~\cite{van2010cross} and CSP~\cite{stamm2010reining} execution of scripts is sandboxed to the specific domain. Nevertheless, there have been attacks that demonstrated that scripts can utilise side channels to read memory or even change content of memory \cite{oren2015spy,gruss2016rowhammer,shusterman2019robust}. Furthermore, malicious scripts do not necessarily need to break out to cause harm, as they have access to the CPU and may utilise this computing power for cryptojacking \cite{eskandari2018first}. Prior Javascript attacks exploited bugs in browser's implementations of SOP \cite{chen2007analysis,barth2009cross,jia2016web,rogowski2017revisiting}.
We demonstrate attacks which do not exploit a bug in the SOP, but combine transport layer attacks in tandem with browser's cache infection to bypass SOP restrictions. This makes our attacks much more difficult to block since they exploit the correct behaviour of the protocols.

There has been work on how to circumvent SOP and CSP by loading images, that contain JavaScript, e.g., \cite{magazinius2013polyglots}. An attacker would still need access to the domain as his image has to be loaded  there, to unleash the full potential of JavaScript. Same applies to the SOP. 
In contrast, our attack infects the domain directly and supplies new methods to create communication channels outside the SOP. The CSP is able to reduce the impact, but as the scripts are executed within the domain, a wide range of attacks may be executed.

\textbf{Browser cache poisoning}. There has been research on how to poison the cache of a browser in different ways, e.g., \cite{bursztein2010bad, vallentin2010persistent,chen2018off}. These show, that caching is a problem at the clientside as the server is not contacted once an element has been cached. This is extended by even injecting malicious code into objects loaded via HTTPS due to users ignoring the certificate errors or supplying malicious browser extensions for the clientside. In this work we show how to create a persistent network of malicious files without a user browsing these while being exposed to an attacker.

Finally, we perform the first experimental study of the attacks that parasite scripts can launch against different applications. We also evaluate the infection, the propagation and the communication of the parasites on popular browsers.

\section{Attacker Model}\label{sc:attack}
Our attacker consists of a master which injects malicious scripts - we call {\em parasites} - into the browser caches of the victim clients. The master then controls the parasites and uses them to launch sophisticated attacks against the victim clients.

{Master} can eavesdrop on the packets exchanged by the victim client but cannot block or modify them. Such an attacker can be another client on a public Wireless network. The master sees the TCP source port and the TCP sequence number in the segments sent by the client and hence can craft correct response segments impersonating the server, without the need to guess these parameters. Our attack proceeds by injecting TCP segments into the TCP connection between the victim client and the target website.

{Parasites} are scripts coming from the legitimate website modified by the attacker to include a malicious behaviour. The parasite is injected by the master into the communication between the client and the website. It is then cached by the victim client's browser. We show how to construct the parasite so that it stays in the browser persistently over long time periods, even when the victim client shuts down its device or changes network. The parasite propagates and infects multiple other domains in the victim's browser's cache. The parasites communicate between themselves and the master and execute commands and attacks on behalf of the master.

\section{Eviction from Browser's Cache}\label{sec:http-eviction}

To infect an object from some website with a malicious script of the attacker, the attacker has to cause the client to issue an HTTP request for that object. Typically objects from popular websites will be cached in victims' browsers, hence during repeated visits to a website the client will not issue requests for objects that are already cached. As a result the attacker cannot inject its malicious script.

\begin{figure}[!htbp]
\vspace{-10pt}
\centering
\begin{adjustbox}{width=.33\textwidth}
\begin{tikzpicture}[scale=0.66]
\newcommand{\posXV}{0}
\newcommand{\posXA}{5}
\newcommand{\posXS}{10}
\newcommand{\dYSmall}{-0.42}
\newcommand{\dYBig}{-0.6}

% 1: Victim GETs any object
\newcommand{\ofsA}{0}
\filldraw[color=blue!10] (\posXV,\ofsA-1) rectangle (\posXS,\ofsA-0.7);
\draw[->, arrows={-Triangle}, thick] (\posXV,\ofsA-1) -- (\posXA,\ofsA-1) -- (\posXS,\ofsA-1);
\node[] at (\posXA,\ofsA-0.85) {\scriptsize{\texttt{GET any.com}}};
\node[xshift=0.05em,state,inner sep=1pt, minimum size=0pt,anchor=west] at (\posXV,\ofsA-0.85) {\tiny{1}};

% 2.1 Attacker injects cache eviction code
\newcommand{\ofsB}{\ofsA+\dYBig}
\filldraw[color=red!10] (\posXV,\ofsB-1) rectangle (\posXA,\ofsB-0.7);
\draw[->, arrows={-Triangle}, thick] (\posXA,\ofsB-1) -- (\posXV,\ofsB-1);
\node[text width=4cm,align=center] at ({(\posXV+\posXA)/2},\ofsB-0.85) {\scriptsize{\texttt{tcp injection}}};
\node[xshift=-0.05em,state,inner sep=1pt, minimum size=0pt,anchor=east] at (\posXA,\ofsB-0.85) {\tiny{2}};

% 2.2 Victim ignores benign (duplicate/out-of-sequence) response
\newcommand{\ofsC}{\ofsB-0.42}
\filldraw[color=blue!10] (\posXV,\ofsC-1) rectangle (\posXS,\ofsC-0.7);
\draw[->, arrows={-Triangle}, thick] (\posXS,\ofsC-1) -- (\posXV,\ofsC-1);
\node[text width=6cm,align=center] at (\posXA,\ofsC-0.85) {\scriptsize{\texttt{ignored benign response}}};

% 3.1: Victim GETs and caches junk objects
\newcommand{\ofsF}{\ofsC+\dYBig}
\filldraw[color=red!10] (\posXV,\ofsF-1) rectangle (\posXS,\ofsF-0.7);
\draw[->, arrows={-Triangle}, thick] (\posXV,\ofsF-1) -- (\posXA,\ofsF-1) -- (\posXS,\ofsF-1);
\node[] at (\posXA,\ofsF-0.85) {\scriptsize{\texttt{GET attacker.com/junk01.jpg}}};
\node[xshift=0.05em,state,inner sep=1pt, minimum size=0pt,anchor=west] at (\posXV,\ofsF-0.85){\tiny{3}};

% 3.2: Victim GETs and caches junk objects
\newcommand{\ofsG}{\ofsF+\dYSmall}
\filldraw[color=red!10] (\posXV,\ofsG-1) rectangle (\posXS,\ofsG-0.7);
\draw[->, arrows={-Triangle}, thick] (\posXV,\ofsG-1) -- (\posXA,\ofsG-1) -- (\posXS,\ofsG-1);
\node[] at (\posXA,\ofsG-0.85) {\scriptsize{\texttt{GET attacker.com/junk02.jpg}}};
\node[xshift=0.05em,state,inner sep=1pt, minimum size=0pt,anchor=west] at (\posXV,\ofsG-0.85){\tiny{3}};

% 3.3: Victim GETs and caches junk objects
\newcommand{\ofsH}{\ofsG+\dYSmall}
\filldraw[color=red!10] (\posXV,\ofsH-1) rectangle (\posXS,\ofsH-0.7);
\draw[->, arrows={-Triangle}, thick] (\posXV,\ofsH-1) -- (\posXA,\ofsH-1) -- (\posXS,\ofsH-1);
\node[] at (\posXA,\ofsH-0.85) {\scriptsize{\texttt{GET attacker.com/junk03.jpg}}};
\node[xshift=0.05em,state,inner sep=1pt, minimum size=0pt,anchor=west] at (\posXV,\ofsH-0.85){\tiny{3}};

% 3.4: GET ...
\newcommand{\ofsI}{\ofsH+\dYSmall}
\filldraw[color=red!10] (\posXV,\ofsI-1) rectangle (\posXS,\ofsI-0.7);
\draw[->, arrows={-Triangle}, thick] (\posXV,\ofsI-1) -- (\posXA,\ofsI-1) -- (\posXS,\ofsI-1);
\node[] at (\posXA,\ofsI-0.85) {\scriptsize{\texttt{GET ...}}};
\node[xshift=0.05em,state,inner sep=1pt, minimum size=0pt,anchor=west] at (\posXV,\ofsI-0.85){\tiny{3}};

% 3.5: GET popular
\newcommand{\ofsJ}{\ofsI+\dYSmall}
\filldraw[color=red!10] (\posXV,\ofsJ-1) rectangle (\posXS,\ofsJ-0.7);
\draw[->, arrows={-Triangle}, thick] (\posXV,\ofsJ-1) -- (\posXA,\ofsJ-1) -- (\posXS,\ofsJ-1);
\node[] at (\posXA,\ofsJ-0.85) {\scriptsize{\texttt{GET popular.com/img.png}}};
\node[xshift=0.05em,state,inner sep=1pt, minimum size=0pt,anchor=west] at (\posXV,\ofsJ-0.85){\tiny{3}};

% 0: Victim entity Line and Label
\newcommand{\ofsZ}{\ofsJ+\dYSmall-1}
\draw[->, arrows={-Triangle}] (\posXV,-0.5) -- (\posXV,\ofsZ);
\node[anchor=north] at (\posXV+0.45,0) {Victim};

% 0: Attacker entity line and label
\begin{scope}[on background layer]
    \draw[->, arrows={-Triangle}] (\posXA,-0.5) -- (\posXA,\ofsZ);
    \node[anchor=north] at (\posXA,0) {Attacker};
\end{scope}

% 0: Server entity line and label
\draw[->, arrows={-Triangle}] (\posXS,-0.5) -- (\posXS,\ofsZ);
\node[anchor=north] at (\posXS-0.45, 0) {Server};

\end{tikzpicture}
\end{adjustbox}
  \vspace{-10pt}
  \caption{Cache eviction. Legitimate messages are blue, attack are red.}
  \vspace{-8pt}
  \label{fig:cache_eviction}
\end{figure}
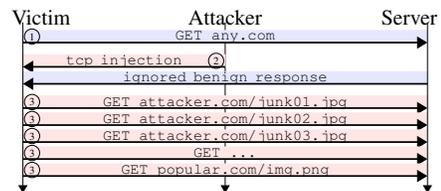

To force the browsers to retrieve the newest copy of the object we perform an enforced cache cleaning. We developed a {\em cache eviction} module, which removes (evicts) all the objects from the browser's cache. Cache cleaning is illustrated in Figure~\ref{fig:cache_eviction} and proceeds as follows: When a user establishes a connection to some website, the attacker injects a spoofed TCP segment (a small inline script), impersonating an authentic TCP segment from the website, into a TCP connection between the user and the website. The injection is illustrated in Figure~\ref{fig:cache_eviction}~\circlenumber{2}.
This script dynamically loads  requests for junk objects (images) in the attacker's domain. The responses populate the cache with the objects and supplant (older) cached elements. The {\em cache eviction} procedure ensures that all further requests for objects generate new retrieval requests from the web server (Figure~\ref{fig:cache_eviction}~\circlenumber{3}), which responses get modified by the attacker in the next step.

{\bf Evaluation.} We evaluated cache eviction against popular browsers\footnote{\url{https://www.w3schools.com/browsers/}}. The browsers differ in cache types, sizes and cache eviction methodologies. We also investigated, whether they allocate memory per domain. If cache capacity is shared between all domains, even if a victim cached objects of \texttt{a.com}, they may be evicted upon visit to \texttt{b.com}, for example. The results are shown in Table~\ref{tab:A}.
\begin{table}[!htbp]\renewcommand{\arraystretch}{0.8}
{\scriptsize{
\begin{tabular}{|c|c|c|c|c|c|c|}
\cline{1-6}
\textbf{Browser}&\textbf{Version}&\textbf{Ev.}&\textbf{I.D.}&\textbf{Size}&\textbf{Remarks}\\
\cline{1-6}
Chrome&81.0.4044.122&\cmark&\cmark& 320MiB\textsuperscript{$\dagger$} & \textsuperscript{$\dagger$}from Chromium \\\hline
Chrome*&81.0.4044.122&\cmark&\cmark& &*incognito mode\\\hline
Edge&84.0.522.59&\hspace{-0.15em}\cmark&\hspace{-0.15em}\cmark& 320MiB\textsuperscript{$\dagger$} & \\\hline
IE&11.1365.17134.0&\hspace{-0.15em}\xmark&\hspace{-0.15em}\xmark&330MB&DOS on memory\\\hline
Firefox&75.0&\cmark&\cmark& 256MB &performance impact\\\hline
Opera&68.0.3618.56&\cmark&\cmark& 320MiB\textsuperscript{$\dagger$}& \\
\cline{1-6}
\end{tabular}
\caption{{\scriptsize{Evaluation of cache eviction on popular browsers. `Ev.' is eviction, `I.D.' is inter-domain, and `Size' represents default cache size.}}}}}
\vspace{-15pt}
\label{tab:A}
\label{tab:cache-eviction}
\end{table}
Eviction from Chromium-based browsers and Firefox is performed easily and efficiently. It has been observed, that Firefox, while evicting cache, may experience reduced responsiveness due to overloaded memory and disk cache. Internet Explorer behaves differently: it appears to allocate more and more space to the memory until the operating system shuts down processes due to low free memory.

\section{Injection of Parasites into TCP Connection}\label{sc:inject:transport}
Injection of scripts can be done with an off-path attacker that is not located on the same network as the victim, e.g.,  via DNS cache poisoning or BGP prefix hijacking \cite{brandt2018domain,birge2018bamboozling}, or via injection of TCP segments by inferring the ACK number and sequence number (SN) \cite{chen2018off}. 
However, since the focus of our work is on evaluating the attacks that such scripts can launch against different applications when running on popular browsers, we perform the injection of script in an eavesdropping attacker mode into the browsers. We assume that our eavesdropping attacker is located on the same wireless network as the victim, e.g., on a public WiFi network.

\begin{table}[!htbp]
\renewcommand{\arraystretch}{0.8}
{\scriptsize
\centering
\begin{tabular}{|c|c|c|c|c|c|c|}
\cline{1-7} \textbf{OS}&\textbf{Chrome}&\textbf{Firefox}&\textbf{IE}&\textbf{Edge}&\textbf{Safari}&\textbf{Opera} \\\hline
{Win10}&\cmark&\cmark&\cmark&\cmark&\cmark&\cmark\\\hline
{MacOS}&\cmark&\cmark&n/a&n/a&\cmark&\cmark\\\hline
{Linux}&\cmark&\cmark&n/a&n/a&n/a&\cmark\\\hline
{Android}&\cmark&\cmark&n/a&n/a&n/a&\cmark\\\hline
{iOS}&\cmark&\cmark&n/a&n/a&\cmark&\cmark\\
\hline
\end{tabular}
\caption{{\scriptsize TCP Injection evaluation. No support by an OS marked 'n/a'.}}\label{tab:tcp_injection}}
\vspace{-15pt}
\end{table}

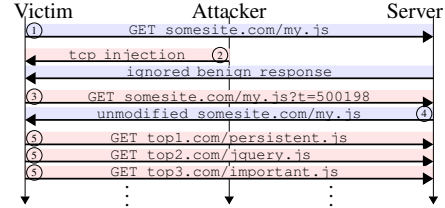
\begin{figure}[!htbp]
\vspace{-15pt}
\centering
\begin{adjustbox}{width=.333\textwidth}
\begin{tikzpicture}[scale=.66]
\newcommand{\posXV}{0}
\newcommand{\posXA}{5}
\newcommand{\posXS}{10}
\newcommand{\dYSmall}{-0.42}
\newcommand{\dYBig}{-0.6}

% 1: Victim GETs target object
\newcommand{\ofsA}{0}
\filldraw[color=blue!10] (\posXV,\ofsA-1) rectangle (\posXS,\ofsA-0.7);
\draw[->, arrows={-Triangle}, thick] (\posXV,\ofsA-1) -- (\posXA,\ofsA-1) -- (\posXS,\ofsA-1);
\node[] at (\posXA,\ofsA-0.85) {\scriptsize{\texttt{GET somesite.com/my.js}}};
\node[xshift=0.05em,state,inner sep=1pt, minimum size=0pt,anchor=west] at (\posXV,\ofsA-0.85) {\tiny{1}};

% 2.1 Attacker injects parasite
\newcommand{\ofsB}{\ofsA+\dYBig}
\filldraw[color=red!10] (\posXV,\ofsB-1) rectangle (\posXA,\ofsB-0.7);
\draw[->, arrows={-Triangle}, thick] (\posXA,\ofsB-1) -- (\posXV,\ofsB-1);
\node[text width=4cm,align=center] at ({(\posXV+\posXA)/2},\ofsB-0.85) {\scriptsize{\texttt{tcp injection}}};
\node[xshift=-0.05em,state,inner sep=1pt, minimum size=0pt,anchor=east] at (\posXA,\ofsB-0.85) {\tiny{2}};

% 2.2 Victim ignores benign (duplicate/out-of-sequence) response
\newcommand{\ofsC}{\ofsB-0.42}
\filldraw[color=blue!10] (\posXV,\ofsC-1) rectangle (\posXS,\ofsC-0.7);
\draw[->, arrows={-Triangle}, thick] (\posXS,\ofsC-1) -- (\posXV,\ofsC-1);
\node[text width=6cm,align=center] at (\posXA,\ofsC-0.85) {\scriptsize{\texttt{ignored benign response}}};

% 3: Parasite makes victim GET unmodified object with altered cache key
\newcommand{\ofsD}{\ofsC+\dYBig}
\filldraw[color=red!10] (\posXV,\ofsD-1) rectangle (\posXS,\ofsD-0.7);
\node[] at (\posXA,\ofsD-0.85) {\scriptsize{\texttt{GET somesite.com/my.js?t=500198}}};
\draw[->, arrows={-Triangle}, thick] (\posXV,\ofsD-1) -- (\posXA,\ofsD-1) -- (\posXS,\ofsD-1);
\node[xshift=0.05em,state,inner sep=1pt, minimum size=0pt,anchor=west] at (\posXV,\ofsD-0.85) {\tiny{3}};

% 4: Server responds to GET with unmodified object
\newcommand{\ofsE}{\ofsD+\dYSmall}
\filldraw[color=blue!10] (\posXV,\ofsE-1) rectangle (\posXS,\ofsE-0.7);
\draw[->, arrows={-Triangle}, thick] (\posXS,\ofsE-1) -- (\posXA,\ofsE-1) -- (\posXV,\ofsE-1);
\node[] at (\posXA,\ofsE-0.85) {\scriptsize{\texttt{unmodified somesite.com/my.js}}};
\node[xshift=-0.05em,state,inner sep=1pt, minimum size=0pt,anchor=east] at (\posXS,\ofsE-0.85) {\tiny{4}};

% 5.1: Infected victim GETs other (top1) target object to be infested
\newcommand{\ofsF}{\ofsE+\dYBig}
\filldraw[color=red!10] (\posXV,\ofsF-1) rectangle (\posXS,\ofsF-0.7);
\draw[->, arrows={-Triangle}, thick] (\posXV,\ofsF-1) -- (\posXA,\ofsF-1) -- (\posXS,\ofsF-1);
\node[] at (\posXA,\ofsF-0.85) {\scriptsize{\texttt{GET top1.com/persistent.js}}};
\node[xshift=0.05em,state,inner sep=1pt, minimum size=0pt,anchor=west] at (\posXV,\ofsF-0.85){\tiny{5}};

% 5.2: Infected victim GETs other (top2) target object to be infested
\newcommand{\ofsG}{\ofsF+\dYSmall}
\filldraw[color=red!10] (\posXV,\ofsG-1) rectangle (\posXS,\ofsG-0.7);
\draw[->, arrows={-Triangle}, thick] (\posXV,\ofsG-1) -- (\posXA,\ofsG-1) -- (\posXS,\ofsG-1);
\node[] at (\posXA,\ofsG-0.85) {\scriptsize{\texttt{GET top2.com/jquery.js~~~~~}}};
\node[xshift=0.05em,state,inner sep=1pt, minimum size=0pt,anchor=west] at (\posXV,\ofsG-0.85){\tiny{5}};

% 5.3: Infected victim GETs other (top3) target object to be infested
\newcommand{\ofsH}{\ofsG+\dYSmall}
\filldraw[color=red!10] (\posXV,\ofsH-1) rectangle (\posXS,\ofsH-0.7);
\draw[->, arrows={-Triangle}, thick] (\posXV,\ofsH-1) -- (\posXA,\ofsH-1) -- (\posXS,\ofsH-1);
\node[] at (\posXA,\ofsH-0.85) {\scriptsize{\texttt{GET top3.com/important.js~~}}};
\node[xshift=0.05em,state,inner sep=1pt, minimum size=0pt,anchor=west] at (\posXV,\ofsH-0.85){\tiny{5}};

% 5.4: etc.pp.
\newcommand{\ofsI}{\ofsH-0.85+\dYSmall}
\node[] at ({(\posXV+\posXA)/2},\ofsI) {$\vdots$};
\node[] at ({(\posXA+\posXS)/2},\ofsI) {$\vdots$};

% 0: Victim entity Line and Label
\newcommand{\ofsZ}{\ofsI-0.36}
\draw[->, arrows={-Triangle}] (\posXV,-0.5) -- (\posXV,\ofsZ);
\node[anchor=north] at (\posXV+0.45,0) {Victim};

% 0: Attacker entity line and label
\begin{scope}[on background layer]
    \draw[->, arrows={-Triangle}] (\posXA,-0.5) -- (\posXA,\ofsZ);
    \node[anchor=north] at (\posXA,0) {Attacker};
\end{scope}

% 0: Server entity line and label
\draw[->, arrows={-Triangle}] (\posXS,-0.5) -- (\posXS,\ofsZ);
\node[anchor=north] at (\posXS-0.45, 0) {Server};

\end{tikzpicture}
\end{adjustbox}
\vspace{-10pt}
\caption{Cache infection attack. The blue packets indicate responses from the genuine webserver, red packets are those injected by the attacker.}
\label{fig:attackoverview}
\end{figure}

{\bf Setup.} The master monitors the communication on the network, waiting for an HTTP request to one of the objects he has prepared in advance, and injects a TCP segment containing the malicious response once such HTTP request is detected. To ensure that the TCP segment is accepted and correctly reassembled, the master must set the correct TCP destination port, TCP sequence number (SN) and offsets - these fields he can adjust from the HTTP request packets that the victim client sends. We implemented and evaluated injection of TCP segments in communication of popular browsers, and confirmed the attack to be effective, independent of the browser and OS; the results are listed in Table \ref{tab:tcp_injection}. The messages exchange diagram for injecting a TCP segment is illustrated in Figure \ref{fig:attackoverview}.

{\bf Attack.} In step \circlenumber{1} the victim sends a request for \textit{my.js} file. The attacker sends a
response containing a malicious script (a parasite) in step \circlenumber{2}. The parasite contains the same name as the original script that was requested by the client. The functionality of the webpage may be modified since the authentic script was not loaded - this may be detected by the victim. To avoid detection, the parasite (i.e., the newly infected script) issues a request to the website to load the original script using a different URL (the with an ignored request parameter), steps \circlenumber{3} and \circlenumber{4}.
This request satisfies the Same Origin Policy (SOP) limitation and is allowed by the browsers. 
The parasite subsequently initiates propagation to other webpages and issues requests to popular webpages, in steps \circlenumber{5} and on, which in turn get infected by the attacker as in step \circlenumber{1}. This results in multiple parasites, each corresponding to an infected script from one popular webpage. This guarantees availability of this puppet - namely, every time any of these popular websites are accessed, the parasite (corresponding to the requested website) is invoked.

{\bf Discussion.} One of the main countermeasures against TCP injection attacks is to employ encryption. Although this fact has been known for a long period of time, our measurement study found that $21\%$ of the 100,000-top Alexa websites do not use HTTPs and almost $7\%$ of the websites use vulnerable SSL versions (SSL2.0 and SSL3.0). Furthermore, even websites supporting SSL/TLS for communication may be compromised. Recent works demonstrated that off-path attackers can trick Certificate Authorities (CAs) into issuing fraudulent certificates, \cite{birge2018bamboozling,brandt2018domain}. If our attacker uses a fraudulent certificate for some target domain it can similarly inject spoofed TCP segments into communication with that domain.

Often even simpler attacks suffice. We evaluated the 15K-top Alexa domains and found that from the $13\,419$ HTTP(S) responders $67.92\%$ did not provide HSTS headers at all, and only $545$ were contained in Chrome's HSTS preload list, leaving up to $96.59\%$ of the domains vulnerable to SSL stripping attacks.

\section{The Parasite Design and Implementation}\label{sc:parasite}
In this section we show how to select scripts that should be infected with parasites, how to ensure they stay persistently in victim browser cache, how to develop methodologies to allow parasites to propagate between different devices and domains and how to develop a Command and Control channel to communicate between the parasites and the attacker.

\subsection{Infecting Objects with Parasites} The attacker's goal is to select a script from some legitimate domain and to infect it with a parasite. The objects on websites can be changed, renamed or even removed. In that case, the control over the parasite-script instance is lost since it will never be invoked and will eventually be removed from the cache. The goal of the attacker is therefore to select such a script that will ensure long term control over the parasite in the browser of the victim client. We achieve this using two key observations: we select objects which do not change often and in those objects we set the headers so that the cache keeps the injected script for the longest possible time duration.

{\bf Selecting persistent scripts.} Which script should we infect in order to guarantee persistency? Ideally the attacker would search for scripts that do not change often and whose names are stable over long time periods. 
\begin{figure}[!htbp]
\begin{center}
\vspace{-5pt}
\includegraphics[width=0.34\textwidth]{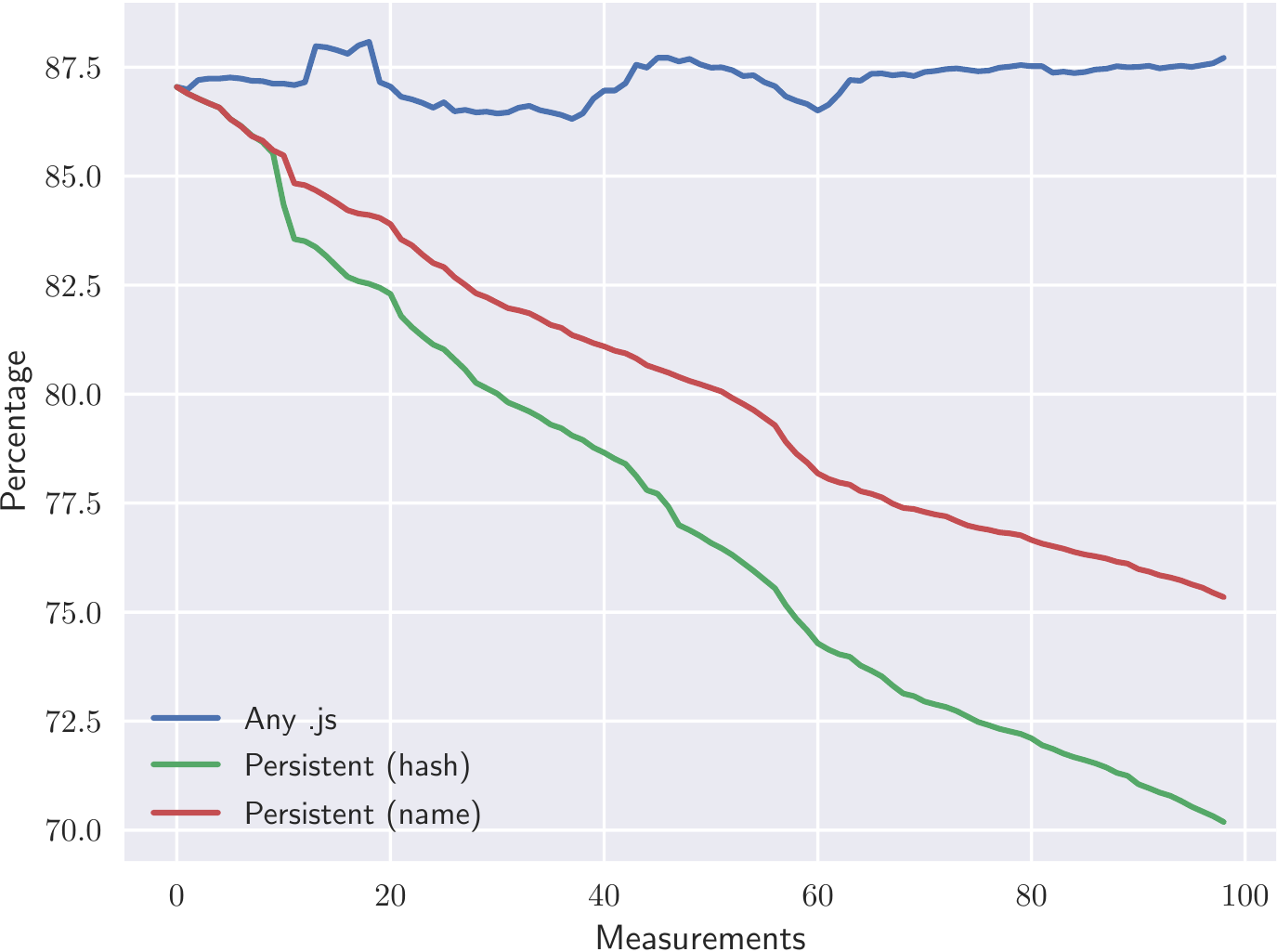}
    \vspace{-10pt}
  \caption{Persistency measurement over 100 days.}\label{fig:persist}
  \vspace{-15pt}
  \end{center}
\end{figure}
To identify such scripts we develop a web crawler to collect statistics over 15K-top Alexa pages. For all objects on these pages, we collect hashes over the files and names, and store them. The web crawler ran daily over a period of 100 days. At the completion, we perform an analysis over the collected data, the results are plotted in Figure \ref{fig:persist}. For instance, in Figure~\ref{fig:persist} for a window of five days about 87.5\% of the websites use at least one persistent object (excluding inline scripts), namely, object that is not renamed over a period of five days. After approximately 100 days, 75.3\% websites are using at least one persistent JavaScript file, which has not been renamed. 

These scripts are perfect targets to be infected with parasites for our persistent botnet, as they are accessed frequently due to the popularity of the websites from which they are served. Figure \ref{fig:persist} also shows that JavaScript files might change in content, while not necessarily in name. If the name was changed within the website, it is not usable for our attack anymore as browsers' caches use names of files as keys and hence, we focus on the name based persistency factor. Using these statistics we are able to select files which remain persistent over time. We use these files as the potential targets for camouflaging our parasites.

{\bf Setting parasite caching headers.} In the JavaScript of the parasite the caching related headers are set to ensure that the browser of the victim keeps the modified copy of the object as long as possible in the cache; the cache duration is set by HTTP headers like the {\tt Cache-Control} header. 

{\bf Infecting scripts with parasites.} When identifying a request for a persistent object in one of the domains of interest the attacker injects a TCP segment containing the original objects attaching the parasite script to the end. 

The attacker loads the original object that would normally be included in the target page. Then the parasite object is created by expanding the original file. "; PARASITE\_CODE;" is appended to the end of the corresponding original JavaScript file. 
For HTML files, a "\textless script\textgreater PARASITE\_CODE\textless /script\textgreater" tag is inserted before the closing "\textless/body\textgreater" tag. However, inserting it in the HTML file is optional so as not to violate any Content Security Policy. The variables and function identifiers in the parasite code have been chosen so that there is no conflict with the target applications.

Afterwards, every time a client makes a request to the server checking for ``freshness'' of the infected object, the request is manipulated to ensure that infected object is refreshed. 

{\bf Requesting the infected objects.} Every time the infected object is requested by the website our parasite script is invoked instead of the original script. The client's request is manipulated. Headers are set which signal to the server that the client has not cached any data. This prevents the server from responding with a 304 status code.
The manipulated request is forwarded to the server. The server's response is also manipulated. If the response is an HTML or JavaScript file, the malware is injected. The original function is preserved by attaching it to the end. We also perform validity checks on the server. The cache headers are adapted, so that the data gets in the cache of the client and remains there for as long as possible. In addition, security headers are removed. This makes it possible to cross-infect other domains. The client caches the parasite and it is executed every time the corresponding resource is loaded. Furthermore all HTTP caches between attacker and victim will be poisoned with the manipulated JavaScript files.

\subsection{Propagation}\label{sc:eval:caches} We developed two methods of propagation for the parasites: the parasites can propagate to other domains on the same victim device and the parasites can propagate to other devices.

\subsubsection{Propagation between domains} After infecting an object from one domain, the parasite can propagate to infecting objects from other domains on the same victim browser. In this section we provide two methods of propagation between domains that we evaluated.

{\bf Propagation on the same device via shared files.} For instance Google Analytics which our measurement found to be used by $63\%$ of the 1M-top Alexa domains \cite{web_builtwith_google_analytics}. Infecting this JavaScript file in the browser cache therefore leads to the parasite being executed on a large number of domains. 

{\bf Propagation via iframes.} For propagation via iframes, the parasite loads the target domains via iframes into the DOM. The browser then loads all the resources associated with these domains.These objects are infected with parasites. This is only possible because all corresponding security headers (more about this in Section \ref{sec:countermeasures}) are disabled. This cross-infection is demonstrated in the demo video\footnote{\url{http://52.144.44.214/demo.html}}. In this video we show, how the visit of a well known and popular site on an insecure Network can lead to the infection of other sites like online banking and web mail, that are not even accessed and used during the active attack.

\subsubsection{Propagation between devices} We experimentally measure how the parasites can propagate between caches of different popular devices. Propagation between devices is made possible by shared network caches but can also be done due to vulnerabilities in web services (e.g., via XSS attacks). The principle is to attack victims behind shared caches: when a victim receives an infected object from the server, all the caching proxies on the way cache that manipulated object. After that all other victims using these proxies receive the malicious cached object.  

\begin{table}[htbp]
\renewcommand{\arraystretch}{0.8}
\centering
{\scriptsize
\begin{tabular}{|c|c|c|c|}
\cline{1-4}
                       \textbf{Browser}       & \textbf{Ctrl+F5} & \textbf{clear cache} & \textbf{clear cookies} \\ \hline
\multicolumn{1}{|c|}{Chrome}  & \xmark       & \xmark          & \cmark             \\ \hline
\multicolumn{1}{|c|}{Firefox} & \xmark       & \xmark          & \cmark             \\ \hline
\multicolumn{1}{|c|}{Edge}    & \xmark       & \xmark           & \cmark \\ \hline
\multicolumn{1}{|c|}{Opera} & \xmark       & \xmark          & \cmark             \\ \hline
\multicolumn{1}{|c|}{IE} & n/a       & n/a          & n/a             \\ \hline
\end{tabular}
\vspace{2pt}
\caption{{\scriptsize{Effectiveness of refresh methods to remove objects stored with Cache API (not supported by IE).}}}
\vspace{-15pt}
\label{tab:cache-api-test}
}
\end{table}

\begin{table*}[ht!]
\renewcommand{\arraystretch}{0.8}
{\scriptsize
\begin{adjustbox}{center,width=\textwidth}
\begin{tabular}{|l|c|l|c|c|l|}
\hline
\textbf{Location} & \textbf{Type} & \textbf{Instance} & \textbf{HTTP} & \textbf{HTTPS} & \textbf{Comment} \\
\hline
\multirow{2}{100pt}{Caches on Victim Host - Client-internal Caches} & \multirow{2}{*}{Browser Cache}  & Desktop & \Cmark & \Cmark & \\
\cline{3-5} & & Smartphones \cite{web_caching_on_smartphones} & \Cmark & \Cmark & \\
\cline{1-5}
\multirow{15}{100pt}{Caches on Victim Network - Client-side Cache} & Transparent Proxy & Squid & \Cmark & \cmark & \\
\cline{2-6} & \multirow{5}{*}{Web Filter} & Cisco Web Security Appliances & \Cmark & \cmark & AsyncOS 9.1.1 \\
\cline{3-6} & & McAfee Web Gateway & \Cmark & \cmark & \\
\cline{3-5} & & Citrix NetScaler \cite{docs_citrix_netscaler_ssl} & \Cmark & \ndmark & \\
\cline{3-5} & & Barracuda Web Filter & \Cmark & \xmark & \\
\cline{3-5} & & Blue Coat ProxySG & \Cmark & \xmark & \\
\cline{2-6} & \multirow{5}{*}{Firewall} & Sophos UTM & ~~\Cmark* & \cmark & *community-documented \\
\cline{3-6} & & Fortigate & \cmark & \cmark & \\
\cline{3-5} & & Barracuda F-Series & \cmark & \xmark & \\
\cline{3-6} & & Cisco ASA & ~~\cmark* & \xmark & *via redirect \\
\cline{3-6} & & pfSense & ~~\cmark* & \xmark & *via squid module \\
\cline{2-6} & \multirow{2}{*}{Transport} & Airplanes \cite{airplane_wifi, docs_airbus_fast} & \cmark & \ndmark & \\
\cline{3-5} & & (Cruise) Vessels \cite{web_cachebox_cruise, web_spliethoff_vessel_caches} & \cmark & \ndmark & \\
\cline{1-5}
\multirow{8}{100pt}{Remote Caches - Backbone and Server-Side Caches} & \multirow{4}{*}{\multirowcell{1}{Reverse Proxies \\ HTTP Accelerators}} & CDNs & \cmark & \cmark & \\
\cline{3-6} & & Varnish HTTP Cache & \cmark & ~~\cmark* & \\
\cline{3-5} & & F5 Big-IP WebAccelerator & \cmark & ~~\cmark* & \multirowcell{1}{*when used with\\ separate SSL Offloader}\\
\cline{3-5} & & SiteCelerate  & \cmark & ~~\cmark* & \\
\cline{2-6} & Web Application Firewall & GoDaddy WAF & \cmark & \ndmark & \\
\cline{2-5} & ISP & CacheMara & \cmark & \xmark & \\
\cline{2-5} & \multirow{2}{*}{Mobile Network} & LTE Network\cite{cacheability_analysis_of_http_traffic_in_4g} & \ndmark & \xmark & \\
\cline{3-6} & & 5G Networks \cite{cooperative_content_caching_in_5g} & \ndmark & \xmark & with MEC \\\hline
\end{tabular}
\end{adjustbox}
}
\caption{{\scriptsize{Evaluation of caches in the wild. Caching enabled by default (\Cmark), optional (\cmark), not supported (\xmark) or supported by architecture model but function not publicly documented or implementation-dependent (\ndmark).}}}
\vspace{-20pt}
\label{tab:cache_types}
\end{table*}
We perform evaluation of parasites' injection into popular caches in the Internet and show that the attack has a substantially wider application scope than merely being applicable to end hosts' (clients) browser caches. The caches that we found vulnerable to our attack are listed in Table \ref{tab:cache_types}. The vulnerability in that case is that the browser cache is shared between multiple sites. As a defence some modern browsers offer separate caches for each `calling context'. Network caches (e.g., on ISP or on local network) do not support such an isolation and can therefore be seen as a shared resource hence enable the associated attacks.
Besides side channel attacks, it is also possible to use network caches as a way of infection. If the entry for a client in the cache is infected, it automatically affects all other clients connected to the cache. Since it is a design feature of network caches to minimise resource usage by sharing resources, but no security mechanisms are provided on the protocol side, all network-based HTTP(s) caches are vulnerable to our attacks. To prevent this, an isolation can be applied in the cache per client, which however would harm performance. Fixing the software of the network cache is not trivial. Injection attacks against reverse proxies (e.g., on CDNs) also affect all users of the infected proxy.

In our evaluations, we saw that different types of caches have different persistence strategies. The persistence of the browsers with regard to Cache API\footnote{\url{https://developers.google.com/web/fundamentals/instant-and-offline/web-storage/cache-api}} was experimentally evaluated using a lab validation server. Our evaluations demonstrated that in all the browsers, cleaning up the cache does not suffice to prevent the attacks. In particular, the cookies must also be deleted in order to remove the parasites from the cache; see Table \ref{tab:cache-api-test}.

\subsection{Command \& Control Channel} \label{subsub:command_and_control} 
\begin{figure}[!htbp]
\vspace{-10pt}
\centering
\begin{adjustbox}{width=0.33\textwidth}
\begin{tikzpicture}[scale=0.66]

% X-Positions of Victim, Internet and "Cache" (the arch)
\newcommand{\posXV}{0}
\newcommand{\posXI}{10}
\newcommand{\posXC}{7.5}

% Skips between the communication arrows
\newcommand{\dYSmall}{-0.42}
\newcommand{\dYBig}{-0.6} % not used for space reasons

% 1: Request for infested object [0]
\newcommand{\ofsA}{0}
\filldraw[color=blue!10] (\posXV,\ofsA-1) rectangle (\posXC,\ofsA-0.7);
\draw[->, arrows={-}, thick] (\posXV,\ofsA-1) -- (\posXC,\ofsA-1);
\node[xshift=0.05em,state,inner sep=1pt, minimum size=0pt,anchor=west] at (\posXV,\ofsA-0.85) {\tiny{1}};
\node[] at ({(\posXV+\posXC)/2},\ofsA-0.85) {\scriptsize{\texttt{GET top2.com/jquery.js}}};

% 2: Response from Cache [-1]
\newcommand{\ofsB}{\ofsA+\dYSmall}
\filldraw[color=blue!10] (\posXV,\ofsB-1) rectangle (\posXC,\ofsB-0.7);
\draw[->, arrows={-Triangle}, thick] (\posXC,\ofsB-1) -- (\posXV,\ofsB-1);
\draw[->, arrows={-}, thick] (\posXC,\ofsA-1) edge[looseness=2, bend left=90] (\posXC,\ofsB-1);
\node[xshift=-0.05em,state,inner sep=1pt, minimum size=0pt,anchor=east] at (\posXC,\ofsB-0.85){\tiny{2}};
\node[text width=4cm,align=center] at ({(\posXV+\posXC)/2},\ofsB-0.85) {\scriptsize{\texttt{load from cache}}};

% 3: Loading original file [-2]
\newcommand{\ofsC}{\ofsB+\dYSmall}
\filldraw[color=red!10] (\posXV,\ofsC-1) rectangle (\posXI,\ofsC-0.7); node[anchor=west]{\footnotesize{in cache}} (5,\ofsC);
\node[xshift=0.05em,state,inner sep=1pt, minimum size=0pt,anchor=west] at (\posXV,\ofsC-0.85) {\tiny{3}};
\node[] at (4,\ofsC-0.85) {\scriptsize{\texttt{reload original file}}};
\draw[->, arrows={-Triangle}, thick] (\posXV,\ofsC-1) -- (4,\ofsC-1) -- (\posXI,\ofsC-1);

% 4: establishing C2 connection [-3]
\newcommand{\ofsD}{\ofsC+\dYSmall}
\filldraw[color=red!10] (\posXV,\ofsD-1) rectangle (\posXI,\ofsD-0.7);
\node[xshift=0.05em,state,inner sep=1pt, minimum size=0pt,anchor=west] at (\posXV,\ofsD-0.85) {\tiny{4}};
\node[] at (4,\ofsD-0.85) {\scriptsize{\texttt{establish C\&C connection}}};
\draw[->, arrows={-Triangle}, thick]  (\posXV,\ofsD-1) -- (4,\ofsD-1) -- (\posXI,\ofsD-1);

% Entity line housekeeping
\newcommand{\ofsZ}{\ofsD-1+\dYSmall}
\draw[->, arrows={-Triangle}] (\posXV,-0.5) -- (\posXV,\ofsZ);
\draw[->, arrows={-Triangle}] (\posXI,-0.5) -- (\posXI,\ofsZ);

\node[anchor=north] at (0.45, 0) {Victim};
\node[anchor=north] at (\posXI-0.65, 0){Internet};

\end{tikzpicture}
\end{adjustbox}
\vspace{-15pt}
  \caption{C\&C communication to parasites.}
  \vspace{-8pt}
\label{fig:attackexecution}
\end{figure}
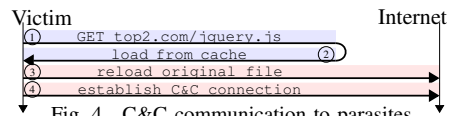

After the victim disconnects from the network on which the initial infection was made, the master uses a C\&C channel to communicate to its parasite instances on the victim. We design and implement a bi-directional C\&C channel for communication between the different parasites on a victim and for communication between the parasites and the remote attacker. Our channel enables sending commands to the bots and retrieving data from the victims. The process is illustrated in Figure \ref{fig:attackexecution}.

Instead of relying on known protocols and features, which can be blocked, such as CORS, we design our own communication protocol. 

We use HTTP information leakage combined with cross requests to allow a remote attacker to create a channel with the parasite scripts in different domains. Our channel is based on the dimensions of a cross image requests. When a cross image request is performed, most image properties are hidden, but the image dimensions are visible. This is needed to adapt the page to the image proportions. For communication we are transferring multiple images, encoding the information in the width and height of the images. Our experiments show that once the dimension is over 65,535, the browsers will downgrade it to this value. Therefore, we can transfer in each image 2 values between 0 and 65,535. Consequently, each image contains 4 bytes of encoded data.
We use {\tt SVG} format, so that the total image size remains small, keeping down the overhead bytes for transferring these 4 bytes of data. An {\tt SVG} image, having no actual content, is of size ~100 bytes, and in our experiments, using a client which sends requests for multiple images simultaneously, we achieve a communication channel of 100KB/s, from the master's server to the parasite.

The communication channel in the other direction, from the parasite to the master's server is as follows: we use HTTP requests, where the URL, or even the URL get/post parameters are the encoded data, and hence with no bandwidth limitations.

\section{Parasite Attacks Against Applications}
\label{sec:applications}
The attacks we developed and evaluated are application dependent. We categorise the attacks per target: against browsers, against operating systems and against networks. The attacks, along with their type, the targets, the exploits and the requirements are listed in Table \ref{tab-victim-caches-effects}. 
We incorporated the different attacks into the parasite via the following modules: (1) a module that reads the browser data such as current URL, user agent, cookies, the local storage; (2) a module for extracting protected browser data, for example, microphone capturing; (3) a module for spreading the parasite based on customised phishing, similar to Emotet \cite{emotet_explanation}; (4) a module that extracts login data, e.g., from Google, Facebook or online banking applications. The parasites use the URL to detect on which website they are currently running, then execute the corresponding attack modules.

\begin{table*}[ht]
\centering
\renewcommand{\arraystretch}{0.618}
{\footnotesize
\begin{adjustbox}{center,width=.8125\textwidth}
\begin{tabular}{|ll|l|l|l|l|}
\cline{1-6} &  & Name & Targets & Exploit & Requirements \\
\hline \multicolumn{1}{|c|}{\multirow{32}{*}{\begin{tabular}[c]{@{}l@{}}Victim \\ Browser\end{tabular}}} & \multirow{5}{*}{C} & Steal Login Data & \begin{tabular}[c]{@{}l@{}}Social networks, web mail, online \\ banking, crypto-exchanges\end{tabular} & \begin{tabular}[c]{@{}l@{}} Use JS access to DOM \& wait for events. Exfiltrate data via C\&C \\ by encoding data to JSON, and send to server with `src' property \\ of an `img' tag that is added to the DOM. We implemented modules \\to read browser data (user agent cookies, local storage), to extract \\ login data by hooking into login forms (e.g., Google, Facebook, online \\ banking apps), and tested them (e.g., on Gmail, Facebook) \end{tabular} & \begin{tabular}[c]{@{}l@{}} $\rightarrow$ if the user is not logged \\ in we wait till he logs in and\\ $\rightarrow$ if the user is logged in \\ we show him a fake login \\ form in the DOM.\end{tabular} \\
\cline{3-6}\multicolumn{1}{|c|}{} &  & Browser Data & Cookies, LocalStorage & Access via Browser API &  no additional requirements\\
\cline{3-6}\multicolumn{1}{|c|}{} &  & Personal Browser Data & Geolocation, microphone, webcam & Access via Browser API & \begin{tabular}[c]{@{}l@{}}Authorization by an \\ attacked domain\end{tabular} \\
\cline{3-6}\multicolumn{1}{|c|}{} &  & Website Data & Financial status, chats, emails... & Access via DOM &  no additional requirements \\
\cline{3-6}\multicolumn{1}{|c|}{} &  & Side Channels & \begin{tabular}[c]{@{}l@{}}Side channels between the \\ browser tabs to communicate \\ on the machine of a victim\end{tabular} & Timing, CPU usage... &  no additional requirements\\
\cline{2-6}\multicolumn{1}{|c|}{} & \multirow{6}{*}{I} & \begin{tabular}[c]{@{}l@{}}Circumvent \\ Two Factor Authentication\end{tabular} & Google Authenticator, TAN... & \begin{tabular}[c]{@{}l@{}}Exploits de-synchronisation of knowledge between server and client.\\Access to the DOM allows attacker to manipulate the data and \\ interfaces the user sees. Attack is done in JS context of attacked site.\end{tabular} & \begin{tabular}[c]{@{}l@{}}No out-of-band transaction \\ detail confirmation is used, \\ or is ignored by the user.\end{tabular} \\
\cline{3-6}\multicolumn{1}{|c|}{} &  & Transaction Manipulation & Online banking, crypto exchanges & \begin{tabular}[c]{@{}l@{}}Let the user think he does his intended transaction, \\ but in reality he will accept an evil transaction\end{tabular} & \begin{tabular}[c]{@{}l@{}}No out-of-band transaction \\ detail confirmation is used, \\ or is ignored by the user.\end{tabular} \\
\cline{3-6}\multicolumn{1}{|c|}{} &  & Send Phishing & \begin{tabular}[c]{@{}l@{}}Web mail, social networks, \\ WhatsApp Web ...\end{tabular} & \begin{tabular}[c]{@{}l@{}}We harvest data out of the chat app, then use the DOM to send \\personalized phishing to the contacts of the user as well as read \\ previous email communication from the DOM.\end{tabular} & \begin{tabular}[c]{@{}l@{}}The application to attack must \\ be open (in a tab) for sending\\ the phishing. It suffices for a\\ browser to be open and used \\ on different sites.\end{tabular} \\
\cline{3-6}\multicolumn{1}{|c|}{} &  & \begin{tabular}[c]{@{}l@{}}Steal Computation \\ Resources\end{tabular} & \begin{tabular}[c]{@{}l@{}}Crypto-currency mining, \\ crack hashes, distributed \\ scraper...\end{tabular} & We use the CPU / GPU to perform computations. &   no additional requirements \\
\cline{3-6}\multicolumn{1}{|c|}{} &  & Click Jacking & Attack noninflected sites & \begin{tabular}[c]{@{}l@{}}We have complete access to the DOM, \\ so we can run click jacking attacks\end{tabular} &  no additional requirements\\
\cline{3-6}\multicolumn{1}{|c|}{} &  & Ad Injection & \begin{tabular}[c]{@{}l@{}}Inject ads in websites \\ the victims visit\end{tabular} & \begin{tabular}[c]{@{}l@{}}We can target revolvers which have many \\ website users on them. Then inject ads in \\ websites the victims visit.\cite{ad_injection_at_scale}\end{tabular} &  no additional requirements\\
\cline{2-6}\multicolumn{1}{|c|}{} & A & DDoS & Attack other sites & \begin{tabular}[c]{@{}l@{}}Use web based requests (images, web sockets...) \\ to overload servers \cite{js_ddos}. An infected network cache, like CDN \\edge server can be exploited for DDoS. \end{tabular} &  no additional requirements\\
\hline\multicolumn{1}{|c|}{\multirow{8}{*}{\begin{tabular}[c]{@{}l@{}}Victim \\ OS\end{tabular}}} & \multirow{3}{*}{C} & JS CPU Cache \& Spectre & Attack the CPU cache via timing & Attacker uses timing side channels to read data in cache \cite{oren2015spy,spectre_js} &  no additional requirements\\
\cline{3-6}\multicolumn{1}{|c|}{} &  & Rowhammer &  Attack the RAM & \begin{tabular}[c]{@{}l@{}}Exploits charges leak of memory cells\\ the exploits use privilege escalation \cite{gruss2016rowhammer} \end{tabular}& \begin{tabular}[c]{@{}l@{}} Lack of HW techniques to \\ prevent the rowhammer \end{tabular}\\
\cline{2-6}\multicolumn{1}{|c|}{} & I & 0-day on Demand & Exploit the System of the client. & \begin{tabular}[c]{@{}l@{}}The parasite loads 0-day exploits to the client and launches them.\end{tabular} & no additional \\
\hline\multicolumn{1}{|c|}{\multirow{3}{*}{\begin{tabular}[c]{@{}l@{}}Victim \\ Network\end{tabular}}} & I & \begin{tabular}[c]{@{}l@{}}Attack Insecure Routers\\ and internal IoT Devices\end{tabular} & \begin{tabular}[c]{@{}l@{}}Attack devices in the internal \\ network of the victim\end{tabular} & \begin{tabular}[c]{@{}l@{}}Use WebRTC and JS to scan and attack \\ devices in the internal network \\ of the victim (sonar.js)\end{tabular} &  no additional requirements\\
\cline{2-6}\multicolumn{1}{|c|}{} & A & DDoS Internal Systems & \begin{tabular}[c]{@{}l@{}}Overload devices in the \\ targeted internal network.\end{tabular} & \begin{tabular}[c]{@{}l@{}}Use infected clients to overload devices \\ in the targeted internal network.\cite{js_ddos}\end{tabular} &  no additional requirements\\
\hline
\end{tabular}
\end{adjustbox}
}
\caption{{\scriptsize{Evaluation of attacks against popular applications, targeting (C)onfidentiality, (I)ntegrity and (A)vailability.}}}
\vspace{-20pt}
\label{tab-victim-caches-effects}
\end{table*}

The vulnerabilities that allow our attacks is the execution of untrusted JS with full access to the DOM. JS has complete read and write access to the DOM, and the submit events can be hooked. By reading data from the DOM the parasite can read email communication, e.g., from Gmail, or account numbers, e.g., on crypto exchanges, or to read the financial status in online banking. 
Encryption of the network traffic does not prevent the attack since the parasite can read the data directly from the input fields and then transfer it to the attacker via C\&C channel. If the user is logged in, a corresponding fake login screen is presented. By manipulating the DOM the attacker can manipulate bank transfer details in online banking. 

Preventing such attacks on the Browser level is hard, because the parasite utilises only standardised JS functions. The most promising way is to limit the communication between the attacker and the parasites by enforcing a strong CSP, Sub-Resource-Integrity and a fresh load of the main HTML file. In this stage of the attack the encryption of the network traffic does not help, because the data is read directly from the DOM and then transferred to the attacker via the Command \& Control methods described in Section \ref{subsub:command_and_control}. 

 The defence to prevent `two-factor authentication' attacks should require the user to confirm the transaction details on a second device. So in addition to the one-time password (OTP) there must be implemented an out-of-band transaction detail confirmation.

 The vulnerabilities that allow advanced attacks, such as phishing, are the same as previously listed. Security-critical applications like web mail should have all the security measures on DNS and HTTP level enabled. Besides CSP and sub-resource-integrity, HSTS should be enabled, because it blocks the attack by enforcing HTTPS. The parasites can also execute side channel attacks against hardware. The defences to prevent such attacks are specific to the low-level systems, the parasites are used only to execute the corresponding JS based exploit code. 
 In addition to attacking the browser and the OS of the victim, we implemented functionality to find other hosts and propagate to them. The parasite uses WebRTC to find the internal IP of the client and runs reconnaissance to find hosts via WebSockets. We fingerprint found hosts by including `img' tags and stylesheets into the DOM, listed to onload events. Once a device is identified, the exploit starts via JS. To execute attacks against victim OS and network the victim has to open any site with manipulated files in the cache.

\vspace{-5pt}
\section{Recommendations for Countermeasures}
\label{sec:countermeasures}

We recommend to disable caching of scripts to ensure that a fresh copy is loaded every time - we implemented this by adding a random query string to each request. The used alphabet in these files is allows to compress the file a lot. 

\begin{figure}[!htbp]\centering
\vspace{-5pt}
\centering
  \includegraphics[width=0.38\textwidth]{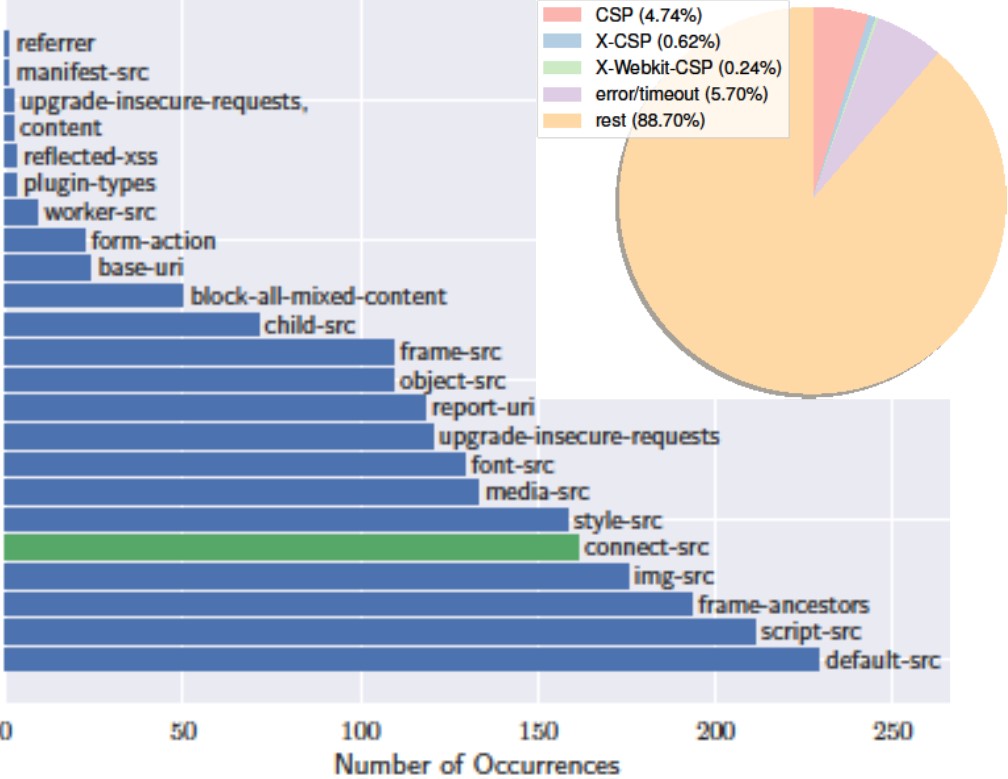}
  \vspace{-10pt}
  \caption{CSP statistics, \texttt{connect-src} frequently used if CSP is supplied. Pie-chart: Used CSP version, where X-CSP and X-Webkit-CSP are deprecated.} \label{fig:cspstats2} \label{fig:cspstats}
  \vspace{-5pt}
\end{figure}

Another defence is cache partitioning which browser vendors started deploying, but studies show that it is inefficient, \cite{rotten_cellar_security_and_privacy_of_the_browser_cache}.
We also recommend that web servers support CSP, which prevents resources from being integrated by third parties. Our measurement of 15K-top Alexa domains shows that CSP is implemented in only 4.33\% of the pages, and only approximately 4.7\% actually supply CSP rules, from which 15.3\% where using a deprecated configuration; see CSP statistics in Figure \ref{fig:cspstats2}. 
Lastly, not well configured headers are supplied in those configurations as well as for example '\texttt{connect-src *;}', which simply allows every \texttt{connect-src} (and therefore also WebSockets without restriction). Out of 160 times '\texttt{connect-src}' being used, 17 used a wildcard here and are therefore not properly configured.

CSP  has  to  be  configured  correctly  to  ensure,  that  only trusted origins are used for remote resources like images  or  WebSockets. 
Even if all the servers are configured correctly (which is not done often~\cite{weichselbaum2016csp}), not all browsers are supporting CSP and there are also bypasses for CSP, which amplifies the problem~\cite{magazinius2013polyglots}.
In order to draw awareness to these headers, browsers should display warnings. Another way to enforce these headers could be that major search engines use them to rank search results, as is done with HTTPS.

It is also recommended to check the integrity of the included resources via the Subresource Integrity (SRI)\footnote{\url{https://developer.mozilla.org/en-US/docs/Web/Security/Subresource_Integrity}} security feature.

However neither CSP nor SRI provide security during the active injection phase by eavesdropping attacker. Since the attacker spoofs the response from the server, it also controls all the headers and the delivered documents. CSP can deliver limited protection when the victim is not exposed to the attacker any more, by eliminating the persistence and the C\&C. 

\vspace{-15pt}
\section{Conclusions}
\label{sec:conclusions}

We develop a botnet which is based on sandboxed scripts, we call parasites, with a remote attacker which controls them. Our methodology of injecting the scripts is based on camouflaging the malicious script as appearing to originate from a genuine website, this allows us to bypass SOP restrictions. Our work demonstrates that the common belief that sandboxed scripts pose an ephemeral threat, which applies only during the visit to the malicious website, is risky. We show that the caches can be forced to store the scripts even after the victim stops visiting the website whose object was infected with a parasite, and even after the victim moves to a different network. We evaluate experimentally the fraction of objects on popular websites that can be exploited for such persistent attacks. 

The main contribution of our work is to experimentally evaluate the attack surface introduced by such parasite scripts on to applications and caches.
\section*{Acknowledgements}
This research has been funded by the German Federal Ministry of Education and Research and the Hessen State Ministry for Higher Education, Research and Arts within their joint support of the National Research Center for Applied Cybersecurity ATHENE, and co-funded by the DFG as part of project S3 within the CRC 1119 CROSSING.

\balance
{\footnotesize 
    \bibliographystyle{splncs04}
    \bibliography{refs,NetSec}
}
\end{document}